\documentclass[aps,prc,notitlepage,%
longbibliography,twocolumn]{revtex4-1}

\usepackage{natbib}
\usepackage{graphicx}
\usepackage{amssymb}
\usepackage{amsmath}
\usepackage{ifthen}
\usepackage{braket}
\usepackage{xcolor}
\usepackage{bm}
\usepackage{bbm}
\usepackage{cancel}
\usepackage{comment}
\usepackage{siunitx}
\usepackage{float}
\usepackage{dcolumn}
\newcolumntype{d}[1]{D{.}{.}{#1}}
\newcolumntype{.}{D{x}{}{9}}
\usepackage{subfigure}
\usepackage{fancyvrb}
\usepackage{maybemath}

\definecolor{garrosgreen}{rgb}{0.1, 0.4, 0.1}
\definecolor{dartmouthgreen}{rgb}{0.05, 0.5, 0.06}
\definecolor{duelferred}{rgb}{0.7, 0.2, 0.1}
\definecolor{cambridgeblue}{rgb}{0.1, 0.3, 1.0}
\definecolor{oxfordblue}{rgb}{0.05, 0.2, 0.7}

\newcommand{\DD}{\mathrm{D}}

\newcommand{\ii}{\mathrm{i}}
\newcommand{\ee}{\mathrm{e}}

\newcommand{\plus}{\mathrm{+}}

\newcommand{\calD}{\mathcal{D}}
\newcommand{\calE}{\mathcal{E}}

\newcommand{\calH}{\mathcal{H}}

\newcommand{\calK}{\mathcal{K}}
\newcommand{\calL}{\mathcal{L}}
\newcommand{\calO}{\mathcal{O}}

\newcommand{\calQ}{\mathcal{Q}}

\newcommand{\FW}{\mathrm{FW}}

\usepackage{xcolor}
\usepackage{bbm}

\begin{document}

\title{Eighth--Order Foldy--Wouthuysen Transformation}

\author{Ulrich D. Jentschura}
\affiliation{Department of Physics and LAMOR, 
Missouri University of Science and
Technology, Rolla, Missouri, USA}

\begin{abstract}
The calculation of higher-order 
binding corrections to bound systems 
is a fundamental problem of theoretical
physics. For any nonrelativistic expansion, 
one needs the Foldy--Wouthuysen Transformation
which disentangles the particle and 
the antiparticle degrees of freedom.
This transformation is carried out here
to eighth order in the momenta, 
or, to eighth order in the momentum operators,
which is equivalent to the eighth order 
of the fine-structure constant.
Matrix elements of the eighth-order terms 
are evaluated for $F_{5/2}$ and $F_{7/2}$ 
states in hydrogenlike ions and compared with 
the Dirac--Coulomb energy levels.
\end{abstract}

\maketitle

\tableofcontents

%
%
\section{Overview}

The Foldy--Wouthuysen transformation~\cite{FoWu1950}
is one of the most essential ingredients of the 
bound-state formalism~\cite{JeAd2022book}. 
Specifically, the Foldy--Wouthuysen transformation
enables one to disentangle the particle and antiparticle 
degrees of freedom, and write 
separate particle and antiparticle
Hamiltonians for spin-$1/2$ particles
coupled to electromagnetic and 
gravitational fields~\cite{JeNo2013pra}.
the case of a free Dirac particle,
the Foldy--Wouthuysen transformation can be 
carried out all orders in the 
momentum operators, and 
the transformed Hamiltonian
takes the simple form
[see Eq.~(11.17) of Ref.~\cite{JeAd2022book}]
\begin{equation}
\label{Hinfty}
H_\FW = \left( \begin{array}{cc}
\sqrt{ \vec p^{\,2} + m^2 } \cdot \mathbbm{1}_{2 \times 2} & 0 \\
0 & -\sqrt{ \vec p^{\,2} + m^2 } \cdot \mathbbm{1}_{2 \times 2} \\
\end{array} \right) \,,
\end{equation}
where $\vec p$ is the momentum operator.
In other cases, where the electron 
is subjected to nontrivial couplings to 
electromagnetic or gravitational fields~\cite{JeNo2013pra}, 
it is possible to carry out the Foldy--Wouthuysen 
transformation only up to a finite order in a 
chosen expansion parameter, which can,
in many cases, be chosen as a typical
momentum scale of the physical problem.
Here, we are concerned with an electron coupled to
a general electromagnetic field, described
by a vector potential $\vec A$ and a 
scalar potential $\Phi$.

For bound systems, including bound Coulomb systems
and electrons bound in a Penning trap~\cite{BrGa1986,WiMoJe2022},
one can identify the typical momentum scale
as $\alpha \, m \, c$, where $m$ is the 
electron mass, $c$ is the speed of light,
and $\alpha$ is the fine-structure constant
or a generalization thereof.
Henceforth, we shall use
natural units with $\hbar = \epsilon_0 = c = 1$,
and $e^2 = 4 \pi \alpha$.
Specifically, for electrons bound in Penning traps, one 
can define a generalized cyclotron 
fine-structure constant according to 
Eq.~(35) of Ref.~\cite{WiMoJe2022}.
We define the kinetic momentum 
\begin{equation} 
\vec \pi = \vec p - e \, \vec A \,,
\end{equation}
where $e$ is the electron charge.
For the electric field $\vec E$ and the 
magnetic field $\vec B$, as well as
its time derivative $e \, \partial_t \vec E$,
we assume the following scaling
[see Eq.~(47) of Ref.~\cite{WiMoJe2022}],
\begin{multline}
\label{scaling}
\vec \pi \sim \alpha \,, \qquad
e \vec A \sim \alpha \,, \qquad
e \, \vec B \sim \alpha^2 \,, 
\\
V \equiv e\, \Phi \,,
\qquad
e \, \vec E = -\vec\nabla V \sim \alpha^3  \,,
\qquad
e \, \partial_t \vec E \sim \alpha^5 \,.
\end{multline}
For the fourth-order Foldy--Wouthuysen
transformation, a particularly
instructive derivation is presented in 
Ref.~\cite{BjDr1964}.
The sixth-order Foldy--Wouthuysen 
transformation has been considered 
extensively in the literature
[see Eqs.~(36)--(38) of Ref.~\cite{ZaPa2010},
Ref.~\cite{HiLePaSo2013}.
Eqs.~(15) and~(20) of Ref.~\cite{PaYePa2016},
Eq.~(7) of Ref.~\cite{HaZhKoKa2020},
as well as Ref.~\cite{ZhMeQi2023}.

One might be surprised about the 
scaling $e \, \vec B \sim \alpha^2$, e.g., when comparing 
to Eq.~(30) of Ref.~\cite{KiNi1996}. Our assumption here is that 
both terms ($\vec p$ and $e \, \vec A$) 
in the relation $\vec\pi = \vec p - e \vec A$
carry the same power of $\alpha$.
Let us consider the vector potential
for a homogeneous magnetic 
trap field in the symmetric gauge, 
$\vec A = \frac12 \, (\vec B \times \vec r)$.
In view of the fact that the position
operator fulfills the scaling 
$|\vec r| \sim \alpha^{-1}$, one requires that 
$e \, \vec B \sim \alpha^2$, to restore the 
scaling of $\vec \pi$. This particular scaling is relevant,
for example, for
the strong field encountered 
in Penning traps~\cite{WiMoJe2022,Je2023mag1,JeMo2023mag2},
where the fine-structure constant
$\alpha$ finds a natural generalization
in terms of a cyclotron fine-structure constant $\alpha_c$.

Here, it is our goal to extend the 
formalism to the eighth order in the 
fine-structure constant. 
We venture to obtain the general Hamiltonian
for electromagnetic coupling
in Sec.~\ref{sec2},
and apply the obtained results to
subsets of bound states in hydrogenlike
systems in Sec.~\ref{sec3},
which leads to a verification of the 
results against the analytically known 
Dirac--Coulomb energy.
Extensive use is made of computer algebra~\cite{Wo1999}.
Conclusions are reserved in Sec.~\ref{sec4}.

%
%
\section{Direct Calculation}
\label{sec2}

\subsection{Unitary Transformation}
\label{sec2A}

We start from the well-known Dirac Hamiltonian $H_\DD$
coupled to a general electromagnetic field,
\begin{align}
\calH_{\rm D} = & \;
\vec \alpha \cdot \vec \pi + \beta \, m 
\nonumber\\
=& \; \left( \begin{array}{cc}
(m + e \, \Phi) \mathbbm{1}_{2 \times 2} & \;\;\; \vec\sigma \cdot \vec \pi \\
\vec\sigma \cdot \vec \pi & \;\;\; (-m + e \, \Phi) \, \mathbbm{1}_{2 \times 2} 
\end{array} \right) \,,
\nonumber\\
\vec\alpha =& \; \left( \begin{array}{cc}
0 & \vec\sigma \\
\vec\sigma & 0
\end{array} \right) \,,
\qquad
\beta = \left( \begin{array}{cc}
\mathbbm{1}_{2 \times 2} & 0_{2 \times 2} \\
0_{2 \times 2} & -\mathbbm{1}_{2 \times 2} 
\end{array} \right) \,.
\end{align}
The Dirac Hamiltonian couples upper and lower components of the Dirac bispinor. 
The aim of the (unitary) Foldy--Wouthuysen transformation is
to separate the upper (particle) from the lower (antiparticle)
degrees of freedom, through an iterative
procedure. One writes the Hamiltonian
as a sum of even terms ($\calE$) and odd terms ($\calO$)
in bispinor space (see also Chap.~11 of Ref.~\cite{JeAd2022book}),
\begin{subequations}
\begin{align}
\calH_{\rm D} =& \; \calE + \calO \,,
\\
\calE =& \;
\tfrac12 \, \left( H + \beta \, H \, \beta \right) \,,
\qquad
\calO =
\tfrac12 \, \left( H - \beta \, H \, \beta \right) \,,
\\
\calE =& \; \beta m + e \Phi
\nonumber\\
=& \; \left( \begin{array}{cc}
(m + e \, \Phi) \mathbbm{1}_{2 \times 2} & \;\;\; 0_{2 \times 2} \\
0_{2 \times 2}  & \;\;\; (-m + e \, \Phi) \, \mathbbm{1}_{2 \times 2} 
\end{array} \right) \,,
\\
\label{defO}
\calO =& \; \vec \alpha \cdot \vec \pi = 
\left( \begin{array}{cc}
0 & \;\;\; \vec\sigma \cdot \vec \pi \\
\vec\sigma \cdot \vec \pi & 0
\end{array} \right) \,.
\end{align}
\end{subequations}
The aim is to eliminate the odd terms $\calO$ through 
unitary transformations. These transformations,
necessarily, in order to preserve
the physical interpretation of the operators,
need to conserve parity
[see the remarks following Eq.~(7.33) of 
Ref.~\cite{ObSiTe2017} and the comprehensive 
discussion in Ref.~\cite{JeNo2014jpa}].
In short, it has been shown 
in Ref.~\cite{JeNo2014jpa} that, if one uses 
a unitary transformation which breaks parity,
the Dirac Hamiltonian can be ``disentangled''
into what seems to be ``particle'' and ``antiparticle''
Hamiltonians, but the operators inside the 
disentangled (diagonal) Hamiltonian have changed
their physical interpretation.
Spurious terms~\cite{Ob2001} which could otherwise break
particle-antiparticle symmetry in gravitational
fields were shown to be absent in 
Ref.~\cite{JeNo2014jpa}, if the parity-conserving,
standard Foldy--Wouthuysen transformation is used
(see also Ref.~\cite{Je2020physics}).
One defines the Hermitian operator $S$ and the 
unitary operator $U = \exp(\ii S)$ as follows,
\begin{equation}
\label{defSandU}
S = -\ii \, \beta \, \frac{\calO}{2 m} \,,
\qquad
S = S^\plus \,,
\qquad
U = \ee^{\ii S} \,.
\end{equation}
The odd operator $\calO$ defined in Eq.~\eqref{defO}
is proportional to the kinetic momentum $\vec \pi$,
and hence, the Foldy--Wouthuysen transformation eliminates
the odd terms order by order in an expansion in the 
momenta. The transformed Hamiltonian is 
written in terms of nested commutators,
\begin{align}
\label{HFWtrafo}
\calH_{\rm FW} =& \; \exp(\ii S) \, \left( H - \ii \, \partial_t \right) 
\exp(-\ii S) 
\nonumber\\[0.1133ex]
=& \; H + \left[ \ii \, S, H - \ii \, \partial_t \right]
+ \frac{1}{2!} \, \left[ \ii S, \left[ \ii \, S,
H  - \ii \, \partial_t \right] \right]
\nonumber\\[0.1133ex]
& \; + \frac{1}{3!} \,
\left[ \ii S, \left[ \ii S, \left[ \ii \, S,
H  - \ii \, \partial_t \right] \right] \right]
\nonumber\\[0.1133ex]
& \; + \frac{1}{4!} \, \left[ \ii S, \left[ \ii S,
\left[ \ii S, \left[ \ii \, S,
H  - \ii \, \partial_t \right] \right] \right] \right] + \ldots \,.
\end{align}
where one notes the identity $[ \ii S, H - \ii \partial_t ] = 
\ii [S, H ] - \partial_t S$.

Though the intensive use of computer algebra 
generalized to the symbolic commutation relations~\cite{Wo1999}
of kinetic momentum operators with the four-vector 
and scalar potentials,
it is possible to carry out the transformation
through eighth order in the fine-structure
constant, under the proviso of the 
scaling implied by Eq.~\eqref{scaling}.
The result of the iterative 
eighth-order transformation~\cite{JeNo2014jpa}, 
can be written as follows,
\begin{equation}
\calH_{\rm FW} = \calH^{[0]} + \calH^{[2]} + \calH^{[4]}
+ \calH^{[6]} + \calH^{[8]} \,.
\end{equation}
The superscript denotes the power of the coupling parameter
at which the term becomes relevant.
The coupling parameter is usually denoted as $\alpha$.
From the zeroth to the third order in $\alpha$, the terms read as follows,
\begin{align}
\label{calH0}
\calH^{[0]} =& \; \beta m \,,
\qquad
\vec \Sigma =
\left( \begin{array}{cc}
\vec\sigma & 0_{2 \times 2} \\
0_{2 \times 2} & \vec \sigma
\end{array} \right) \,,
\\[0.1133ex]
\label{calH2}
\calH^{[2]} =& \; \frac{\beta}{2 m} ( \vec\Sigma \cdot \vec\pi )^2 + V \,.
\end{align}
The $\alpha^4$ terms can be expressed very compactly,
and are found in agreement with Refs.~\cite{BjDr1964,Pa2005},
\begin{align}
\label{calH4}
\calH^{[4]} =& \; 
-\beta \, \frac{1}{8 m^3} ( \vec\Sigma \cdot \vec\pi )^4
- \frac{\ii \, e}{8 m^2} [ \, \vec\Sigma \cdot \vec\pi,  
\vec\Sigma \cdot \vec E \, ] \,.
\end{align}
For the $\alpha^6$ terms, we indicate three 
alternative representations,
\begin{align}
\label{calH6}
\calH^{[6]} =& \; 
\frac{\beta \, ( \vec\Sigma \cdot \vec\pi )^6 }{16 m^5}
- \frac{5 \ii e}{128 m^4} 
[ \, \vec\Sigma \cdot \vec\pi, [ \, \vec\Sigma \cdot \vec\pi, 
[ \, \vec\Sigma \cdot \vec\pi, \vec\Sigma \cdot \vec E \, ] \, ] \, ]
\nonumber\\[0.1133ex]
& \; + \frac{\ii \, e }{8 m^4} \; 
\{ \, (\vec\Sigma \cdot \vec \pi)^2, 
[ \, \vec\Sigma \cdot \vec \pi, \vec\Sigma \cdot \vec E \, ] \}
+ \beta \, \frac{e^2 \vec E^{\,2} }{8 m^3} 
\nonumber\\
=& \; 
\frac{\beta \, ( \vec\Sigma \cdot \vec\pi )^6 }{16 m^5}
+  \frac{5 \ii e}{128 m^4}
\left[ ( \vec\Sigma \cdot \vec\pi )^2 ,
\left\{ \vec\Sigma \cdot \vec E, \vec\Sigma \cdot \vec\pi \right\} \right] 
\nonumber\\[0.1133ex]
& \; + \frac{3 \ii \, e}{64 m^4} \;
\{ \, (\vec\Sigma \cdot \vec \pi)^2,
[ \, \vec\Sigma \cdot \vec \pi, \vec\Sigma \cdot \vec E \, ] \}
+ \beta \, \frac{e^2 \vec E^{\,2} }{8 m^3} \,.
\nonumber\\[0.1133ex]
=& \;
\frac{\beta \, ( \vec\Sigma \cdot \vec\pi )^6 }{16 m^5}
+  \frac{3 \ii e}{32 m^4}
\left[ \left( \vec\Sigma \cdot \vec\pi \right)^3 ,
\vec\Sigma \cdot \vec E \right] \,.
\nonumber\\[0.1133ex]
& \; - \frac{\ii \, e}{128 m^4} \;
[ \, (\vec\Sigma \cdot \vec \pi)^2,
\{ \, \vec\Sigma \cdot \vec \pi, \vec\Sigma \cdot \vec E \, \} \, ]
+ \beta \, \frac{e^2 \vec E^{\,2} }{8 m^3} \,.
\end{align}
Here, $\{ A, B \} = A \, B + B \, A$ denotes the anticommutator.
The last form of the sixth-order terms 
in Eq.~\eqref{calH6} is in agreement with the 
sixth-order terms from Eq.~(8) Ref.~\cite{ZhMeQi2023}.
The sixth-order terms are also compatible with
Eqs.~(36)--(38) of Ref.~\cite{ZaPa2010},
with Eq.~(7) of Ref.~\cite{HaZhKoKa2020},
and with the approach from Ref.~\cite{HiLePaSo2013}.
Alternatively, the result in Eq.~\eqref{H6} can be
obtained by applying the unitary
transformation outlined in Eq.~(19) of Ref.~\cite{PaYePa2016}
to the Hamiltonian given in Eq.~(15) of Ref.~\cite{PaYePa2016},
which is tantamount to the 
Hamiltonian obtained by adding the
terms given in Eqs.~(15) and~(20) of Ref.~\cite{PaYePa2016}.

The eighth-order term are naturally written as a
sum of a kinetic term $\calK$, a term $\calD$
involving temporal derivatives of the electric field,
terms quadratic in the electric field,
denoted by $\calQ$,
and linear terms in the electric field,
which we denote by $\calL$. The result is
as follows,
\begin{widetext}
\begin{align}
\label{calH8}
\calH^{[8]} = & \; \calK + \calD + \calQ + \calL \,,
\qquad
\calK = - \beta \, \frac{5}{128 m^7} ( \vec\Sigma \cdot \vec\pi )^8 \,,
\qquad
\calD = - \frac{\ii \, e^2}{32 m^4} \, [ \,
\vec\Sigma \cdot \vec E,
\vec\Sigma \cdot \partial_t \vec E \, ] 
+ \frac{e}{48 m^5} \, \beta \,
\{ \, (\vec\Sigma \cdot \vec \pi)^3, \vec\Sigma \cdot \partial_t \vec E \, \} \,,
\nonumber\\
\calQ =& \;
\frac{7 \, \beta \, e^2}{192 m^5} \,
[ \, \vec\Sigma \cdot \vec \pi, \vec\Sigma \cdot \vec E \, ] \;
[ \, \vec\Sigma \cdot \vec \pi, \vec\Sigma \cdot \vec E \, ]
- \frac{3 \, \beta \, e^2}{64 m^5} \, 
\{ \, \vec\Sigma \cdot \vec \pi, \vec\Sigma \cdot \vec E \, \} \;
\{ \, \vec\Sigma \cdot \vec \pi, \vec\Sigma \cdot \vec E \, \}
- \frac{\beta e^2 }{24 m^5} \, 
[\, \vec\Sigma \cdot \vec \pi, [ \, \vec\Sigma \cdot \vec \pi,
(\vec\Sigma \cdot \vec E)^2 \, ] \, ] \,,
\nonumber\\
\calL = & \;
- \frac{5 \, \ii \, e}{1024 m^6} 
[ \, \vec\Sigma \cdot \vec \pi, [ \, \vec\Sigma \cdot \vec \pi,
[ \, \vec\Sigma \cdot \vec \pi, [ \, \vec\Sigma \cdot \vec \pi,
[ \, \vec\Sigma \cdot \vec \pi,
\vec\Sigma \cdot \vec E
\, ]  \,] \,] \,] \, ]
- \frac{\ii \, e}{32 m^6}
\{ \, \vec\Sigma \cdot \vec \pi, \{ \, \vec\Sigma \cdot \vec \pi,
[ \, \vec\Sigma \cdot \vec \pi, [ \, \vec\Sigma \cdot \vec \pi,
[ \, \vec\Sigma \cdot \vec \pi,
\vec\Sigma \cdot \vec E
\, ]  \,] \,] \} \}
\nonumber\\
& \; - \frac{\ii \, e}{48 m^6} \,
\{ \, \vec\Sigma \cdot \vec \pi, \{ \, \vec\Sigma \cdot \vec \pi,
\{ \, \vec\Sigma \cdot \vec \pi, \{ \, \vec\Sigma \cdot \vec \pi,
[ \, \vec\Sigma \cdot \vec \pi,
\vec\Sigma \cdot \vec E
\, ]  \, \}  \, \} \, \} \, \} \,.
\end{align}
For alternative representations, one notes the identities 
\begin{subequations}
\label{calH8prime}
\begin{align}
\label{calQalt}
\calQ =& \; -\frac{\beta e^2}{96 m^5} \,  \;
\{ \, \vec\Sigma \cdot \vec E, 
4 \, (\vec\Sigma \cdot \vec \pi)^2 \, \vec\Sigma \cdot \vec E +
4 \, \vec\Sigma \cdot \vec E (\vec\Sigma \cdot \vec \pi)^2 +
\vec\Sigma \cdot \vec \pi \, \vec\Sigma \cdot \vec E \, 
\vec\Sigma \cdot \vec \pi \, \} \,, \\
\label{calLalt}
\calL =& \; \frac{65 \ii e}{3072 m^6} [ (\vec\Sigma \cdot \vec \pi)^4, 
\{ \, \vec\Sigma \cdot \vec \pi, \, \vec\Sigma \cdot \vec E \, \} ] 
- \frac{77 \ii e}{1536 m^6} [ (\vec\Sigma \cdot \vec \pi)^5, 
\vec\Sigma \cdot \vec E ]
\nonumber\\
& \; - \frac{43 \, \ii \, e}{1536 m^6} \,
[ \, (\vec\Sigma \cdot \vec \pi)^3, \, 
(\vec\Sigma \cdot \vec \pi)^2 \, \vec\Sigma \cdot \vec E +
\vec\Sigma \cdot \vec E \, (\vec\Sigma \cdot \vec \pi)^2 +
\vec\Sigma \cdot \vec \pi \, \vec\Sigma \cdot \vec E \, 
\vec\Sigma \cdot \vec \pi \, ] \,.
\end{align}
\end{subequations}
\end{widetext}
The particle-antiparticle symmetry implies that the 
terms are invariant when the
following transformations are simultaneously applied:
{\em (i)} multiplication by 
an overall factor $-1$, {\em (ii)} replacement $\beta \to -\beta$,
{\em (iii)} replacement $\vec\pi \to -\vec\pi$,
and $\partial_t \to -\partial_t$
{\em (iv)} replacement $\vec \Sigma \to -\vec\Sigma$,
and {\em (v)} 
replacements $e \to -e$, $V \to -V$ and $\vec E \to - \vec E$.

%
%
\subsection{General Particle Hamiltonian}
\label{sec32}
\index{Foldy--Wouthuysen Transformation!General Particle Hamiltonian}

The upper left 
$2 \times 2$ submatrix of $\calH_{\rm FW}$
constitutes the particle Hamiltonian,
while the lower left 
$2 \times 2$ submatrix of $\calH_{\rm FW}$
constitutes the antiparticle Hamiltonian.
Here, we concentrate on the particle Hamiltonian.
Formally, the particle Hamiltonian
can be found from the results given in Eqs.~\eqref{calH2}---\eqref{calH8}
under the replacements $\vec\Sigma \to \vec\sigma$,
and $\beta \to \mathbbm{1}_{2 \times 2}$,
The general Foldy--Wouthuysen transformed
particle Hamiltonian $H_{\rm FW}$ under the presence of the 
external electric and magnetic fields is obtained as
\begin{equation}
\label{HFWaufgedroeselt}
H_{\rm FW} = \beta m + H^{[2]} + H^{[4]} + H^{[6]} + H^{[8]} \,.
\end{equation}
One finds
\begin{align}
H^{[2]} =& \;  \frac{1}{2 m} ( \vec\sigma \cdot \vec\pi )^2 + V \,,
\\
H^{[4]} = & \;
- \, \frac{1}{8 m^3} ( \vec\sigma \cdot \vec\pi )^4
- \frac{\ii \, e}{8 m^2} [ \, \vec\sigma \cdot \vec\pi,  \vec\sigma \cdot \vec E \, ] \,,
\\
\label{H6}
H^{[6]} =& \;
\frac{( \vec\sigma \cdot \vec\pi )^6 }{16 m^5} 
+  \frac{5 \ii e}{128 m^4}
\left[ \left( \vec\sigma \cdot \vec\pi \right)^2 ,
\left\{ \vec\sigma \cdot \vec E, \vec\sigma \cdot \vec\pi \right\} \right] 
\nonumber\\[0.1133ex]
& \; + \frac{3 \ii \, e}{64 m^4} \;
\{ \, (\vec\sigma \cdot \vec \pi)^2,
[ \, \vec\sigma \cdot \vec \pi, \vec\sigma \cdot \vec E \, ] \}
+  \, \frac{e^2 \vec E^{\,2} }{8 m^3} \,.
\end{align}
Tn order to fix ideas, we point out that the
expression $( \vec\sigma \cdot \vec\pi )^2 = \vec \pi^2 - e \,
\vec\sigma\cdot \vec B$ contains both the 
orbital as well as the spin coupling 
to the magnetic field. If 
$\vec A = \frac12 (\vec B \times \vec r)$,
and the homogeneous trap field $\vec B$ is directed
along the $z$ axis, one has
$( \vec\sigma \cdot \vec \pi)^2 =
\vec p^{\,2} - e \vec L \cdot \vec B
+ \frac{m^2 \omega_c^2}{4} \,\rho^2
- e \vec\sigma\cdot \vec B$,
where $\omega_c = \frac{|e| \, B}{m}$ 
is the cyclotron frequency,
and $\rho^2 = x^2 + y^2$ is the 
coordinate perpendicular to the axis of the magnetic field.

The eighth-order term comprise
the kinetic term $K$, a term $D$
involving temporal derivatives of the electric field,
terms quadratic in the electric field,
denoted by $Q$,
and linear terms in the electric field, 
which we denote by $L$,
\begin{widetext}
\begin{align}
\label{H8}
H^{[8]} = & \; K + D + Q + L \,,
\qquad
K = -  \, \frac{5}{128 m^7} ( \vec\sigma \cdot \vec\pi )^8 \,,
\qquad
D = - \frac{\ii \, e^2}{32 m^4} \, [ \, \vec\sigma \cdot \vec E,
\vec\sigma \cdot \partial_t \vec E \, ]
+ \frac{e}{48 m^5} \,
\{  (\vec\sigma \cdot \vec \pi)^3, \vec\sigma \cdot \partial_t \vec E \} \,, 
\nonumber\\
Q =& \; \frac{7 \, e^2}{192 m^5} \,  \; 
[ \, \vec\sigma \cdot \vec \pi, \vec\sigma \cdot \vec E \, ] \;
[ \, \vec\sigma \cdot \vec \pi, \vec\sigma \cdot \vec E \, ]
- \frac{3 \, e^2}{64 m^5} \,  \; 
\{ \, \vec\sigma \cdot \vec \pi, \vec\sigma \cdot \vec E \, \} \;
\{ \, \vec\sigma \cdot \vec \pi, \vec\sigma \cdot \vec E \, \}
- \frac{e^2}{24 m^5} \,  \; 
[\, \vec\sigma \cdot \vec \pi, [ \, \vec\sigma \cdot \vec \pi, 
(\vec\sigma \cdot \vec E)^2 \, ] \, ] \,,
\nonumber\\
L = & \; 
- \frac{5 \ii e}{1024 m^6}  
[ \vec\sigma \cdot \vec \pi, [ \, \vec\sigma \cdot \vec \pi,
[ \, \vec\sigma \cdot \vec \pi, [ \, \vec\sigma \cdot \vec \pi,
[ \, \vec\sigma \cdot \vec \pi, \sigma \cdot \vec E \, ]  \,] \,] \,] ]
- \frac{\ii \, e}{32 m^6} \, 
\{ \, \vec\sigma \cdot \vec \pi, \{ \, \vec\sigma \cdot \vec \pi,
[ \, \vec\sigma \cdot \vec \pi, [ \, \vec\sigma \cdot \vec \pi,
[ \, \vec\sigma \cdot \vec \pi, 
\sigma \cdot \vec E 
\, ]  \,] \,] \, \} \, \}
\nonumber\\
& \; - \frac{\ii \, e}{48 m^6} \, 
\{ \, \vec\sigma \cdot \vec \pi, \{ \, \vec\sigma \cdot \vec \pi,
\{ \, \vec\sigma \cdot \vec \pi, \{ \, \vec\sigma \cdot \vec \pi,
[ \, \vec\sigma \cdot \vec \pi, \sigma \cdot \vec E
\, ]  \, \}  \, \} \, \} \, \} \,.
\end{align}
For alternative representations, one notes identities 
analogous to Eqs.~\eqref{calQalt} and~\eqref{calLalt},
\begin{subequations}
\label{H8prime}
\begin{align}
\label{Qalt}
Q =& \; -\frac{e^2}{96 m^5} \,  \;
\{ \, \vec\sigma \cdot \vec E, 
4 \, (\vec\sigma \cdot \vec \pi)^2 \, \vec\sigma \cdot \vec E +
4 \, \vec\sigma \cdot \vec E (\vec\sigma \cdot \vec \pi)^2 +
\vec\sigma \cdot \vec \pi \, \vec\sigma \cdot \vec E \, 
\vec\sigma \cdot \vec \pi \, \} \,, \\
\label{Lalt}
L =& \; \frac{65 \ii e}{3072 m^6} [ (\vec\sigma \cdot \vec \pi)^4, 
\{ \, \vec\sigma \cdot \vec \pi, \, \vec\sigma \cdot \vec E \, \} ] 
- \frac{77 \ii e}{1536 m^6} [ (\vec\sigma \cdot \vec \pi)^5, 
\vec\sigma \cdot \vec E ]
\nonumber\\
& \; - \frac{43 \, \ii \, e}{1536 m^6} \,
[ \, (\vec\sigma \cdot \vec \pi)^3, \, 
(\vec\sigma \cdot \vec \pi)^2 \, \vec\sigma \cdot \vec E +
\vec\sigma \cdot \vec E \, (\vec\sigma \cdot \vec \pi)^2 +
\vec\sigma \cdot \vec \pi \, \vec\sigma \cdot \vec E \, 
\vec\sigma \cdot \vec \pi \, ] \,.
\end{align}
\end{subequations}

\end{widetext}

%
%
\section{Applications}
\label{sec3}

%
%
\subsection{Coulomb Field Coupling}
\label{coul_lead_coupling}
\index{Coulomb Field}

One of the important applications of the 
Hamiltonian~\eqref{HFWaufgedroeselt} concerns Coulombic bound states,
which are relevant to one-electron
ions in the central field of a nucleus 
of charge number $Z$.
In this case, one has the relations
\begin{align}
\label{scene_coul}
e \, A^0 = & \; V = 
-\frac{Z\alpha}{r} \,, 
\qquad
\vec A = \vec 0 \,,
\qquad
\vec B = \vec 0 \,,
\\
\qquad
e \, \vec E =& \; -e \, \vec\nabla A^0 =
- \vec \nabla V \,,
\qquad
\vec \pi = \vec p \,.
\end{align}
The scalar potential and electric field 
are time-independent in this case.
One ends up with the following leading term,
where the subscript $C$ indicates the 
relevance for the Coulomb field,
\begin{equation}
\label{HDC2}
H^{[2]}_C = \frac{\vec p^{\,2}}{2 m} + V 
= \frac{\vec p^{\,2}}{2 m} -\frac{Z\alpha}{r} \,.
\end{equation}
This is the Schr\"{o}dinger--Coulomb Hamiltonian in 
the nonrecoil approximation
(Chap.~~4 of Ref.~\cite{JeAd2022book}).
For the evaluation of the fourth-order corrections, 
one needs the identities
\begin{subequations}
\begin{align}
\label{ushelp1}
\ii e \, [ \vec{\sigma} \cdot \vec p, \vec{\sigma} \cdot \vec{E} ]
=& \; - \vec{\nabla}^2 V -
2\, \vec\sigma \cdot ( \vec\nabla V\times\vec p ) \,,
\\
e \{ \vec\sigma \cdot \vec p, \vec\sigma \cdot \vec E \}
=& \; - \{ \vec\sigma \cdot \vec p, \vec\sigma \cdot \vec\nabla V \}
= - \ii [ \vec p^{\,2}, V ] \,.
\end{align}
\end{subequations}
For the Coulomb field,
the well-known leading relativistic correction to the 
Foldy--Wouthuysen Hamiltonian reads as follows,
\begin{widetext}
\begin{equation}
\label{HDC4}
H^{[4]}_C =
- \frac{\vec p^{\,4}}{8 m^3} 
+ \frac{1}{8 m^2} \vec{\nabla}^2 V 
+ \frac{1}{4 m^2} \vec\sigma \cdot ( \vec\nabla V\times\vec p ) 
=
- \frac{\vec p^{\,4}}{8 m^3} 
+ \frac{\pi \, (Z\alpha)}{2 m^2} \; 
\delta^{(3)}(\vec r) + \frac{Z\alpha}{4 m^2 \, r^3} \; 
\vec\sigma \cdot \vec L \,,
\end{equation}
where $\vec L$ is the orbital angular momentum operator.

The sixth-order corrections attain the following form,
\begin{equation}
\label{HDC6}
H^{[6]}_C = \frac{\vec p^{\,6}}{16 m^5}
- \frac{3
\{ \, \vec p^{\,2}, \vec\nabla^2 V \, \}
}{64 m^4} \;
- \frac{3
\{ \, \vec p^{\,2}, 
\vec \sigma \cdot ( \vec\nabla V \times \vec p ) \} 
}{32 m^4} \;
+  \frac{5}{128 m^4} \,
\left[ \vec p^{\,2}, [ \vec p^{\,2}, V ] \right]
+  \, \frac{(\vec\nabla V)^{\,2} }{8 m^3} \,.
\end{equation}
With $\partial_t \vec E = \vec 0$, one has for the eighth-order 
corrections,
\begin{subequations}
\begin{align}
\label{HDC8}
H^{[8]}_C = & \; K_C + D_C + Q_C + L_C \,,
\qquad
K_C = -  \frac{5}{128 m^7} \vec p^{\,8} \,,
\qquad
D_C = 0 \,,
\\
Q_C = & \;
\frac{7
[ \vec{\nabla}^2 V + 2\, \vec\sigma \cdot ( \vec\nabla V\times\vec p ) ] \, 
[ \vec{\nabla}^2 V + 2\, \vec\sigma \cdot ( \vec\nabla V\times\vec p ) ]
}{192 m^5} 
+ \frac{3
\,  \; [ \vec p^{\,2}, V ] \,  [ \vec p^{\,2}, V ] 
}{64 m^5} 
- \frac{ [\, \vec\sigma \cdot \vec p, [ \, \vec\sigma \cdot \vec p,
(\vec \nabla V)^2 \, ] \, ] }{24 m^5} \,,
\\
L_C =& \;
\frac{5
[ \vec\sigma \cdot \vec p, [ \, \vec\sigma \cdot \vec p,
[ \, \vec\sigma \cdot \vec p, [ \, \vec\sigma \cdot \vec p,
\vec{\nabla}^2 V + 2\, \vec\sigma \cdot ( \vec\nabla V\times\vec p ) 
]  \,] \,] \,] 
}{1024 m^6}
+ \frac{\{ \, \vec\sigma \cdot \vec p, \{ \, \vec\sigma \cdot \vec p,
[ \, \vec\sigma \cdot \vec p, [ \, \vec\sigma \cdot \vec p,
\vec{\nabla}^2 V + 2\, \vec\sigma \cdot ( \vec\nabla V\times\vec p ) 
\,] \,] \, \} \, \}
}{32 m^6} \,
\nonumber\\
& \; + \frac{1}{48 m^6} \,
\{ \vec\sigma \cdot \vec p, \{ \vec\sigma \cdot \vec p,
\{ \vec\sigma \cdot \vec p, \{ \vec\sigma \cdot \vec p,
\vec{\nabla}^2 V + 2\, \vec\sigma \cdot ( \vec\nabla V\times\vec p ) 
\}  \} \} \} \,.
\end{align}
\end{subequations}
\end{widetext}

%
%
\subsection{Application: $\maybebm{F_{5/2}}$ States}
\label{sec3A}

We shall compare the eighth-order
corrections to the bound-state energy 
obtained from Eq.~\eqref{HDC8} to the 
bound-state energies of the Dirac--Coulomb problem.
It is well known that the Dirac--Coulomb
problem can be solved exactly (Chap.~8 
of Ref.~\cite{JeAd2022book}), with the result
\begin{align}
E_\DD =& \; m \, 
\left( 1 + \frac{(Z\alpha)^2}{(n_r + \gamma)^2} \right)^{-\tfrac12} \,,
\nonumber\\[0.1133ex]
n_r =& \; n - j - \tfrac12 \,,
\qquad
\gamma = \sqrt{ (j+1/2)^2 + (Z\alpha)^2 } \,.
\end{align}
For $nF_{5/2}$ states, the presence of 
large spin-orbit coupling implies the emergence 
of nontrivial corrections to the energy
from the corresponding higher-order terms
in Eq.~\eqref{HDC8}. One expands as follows,
\begin{multline}
\label{EDnF52}
E_\DD(nF_{5/2}) = m - \frac{(Z\alpha)^2 m}{2 n^2} 
+ (Z\alpha)^4 m \left( \frac{3}{8 n^4} - \frac{1}{6 n^3} \right)  \\
- (Z\alpha)^6 m \left( 
\frac{1}{216 n^3}+\frac{1}{24 n^4}-\frac{1}{4 n^5}+\frac{5}{16 n^6}
\right) 
\\
+ (Z\alpha)^8 m 
\Bigl( -\frac{1}{3888 n^3}-\frac{1}{432 n^4}-\frac{1}{432 n^5}
\\
+\frac{5}{48 n^6}-\frac{5}{16 n^7}+\frac{35}{128 n^8} \Bigr) 
+ \calO(Z\alpha)^{10} \,.
\end{multline}
Within the Foldy--Wouthuysen method, the eighth-order
terms comprise several effects,
namely, the combined effect of third-order 
perturbative terms $E^{[8]} = E^{[8]}_c$ generated by the 
fourth-order Hamiltonian $H^{[4]}$, the 
mixed fourth- and sixth-order terms $E^{[8]}_m$,
and the diagonal element of eighth order
$E^{[8]}_d$,
\begin{equation}
E^{[8]} = E^{[8]}_c + E^{[8]}_m + E^{[8]}_d \,.
\end{equation}
The terms will be further examined in the following.
Let $G'$ be the reduced Green function,
\begin{equation}
G' = \left( \frac{1}{E_S - H_S} \right)' \,,
\end{equation}
where $E_S$ is the Schr\"{o}dinger--Coulomb 
energy $E_S =  - \frac{(Z\alpha)^2 m}{2 n^2}$
and $H_S = H^{[2]}_C$ is the 
Schr\"{o}dinger--Coulomb Hamiltonian.
Then,
\begin{subequations}
\begin{align}
\label{Ec}
E^{[8]}_c =& \;
\langle H^{[4]} \, G' \, (H^{[4]} -\langle H^{[4]} \rangle) 
\, G' \, H^{[4]} \rangle \,,
\\
\label{Em}
E^{[8]}_m =& \;
2 \langle H^{[4]} \, G' \, H^{[6]} \rangle  \,,
\\
\label{Ed}
E^{[8]}_d =& \; \langle H^{[8]} \rangle 
= \langle K_C \rangle +
\langle Q_C \rangle +
\langle L_C \rangle \,.
\end{align}
\end{subequations}
After lengthy algebra, we obtain the results,
\begin{subequations}
\label{IDnF52}
\begin{align}
\frac{\langle K_C \rangle_{nF_{5/2}}}{(Z\alpha)^8 m} =& \;
-\frac{2}{693 n^3}+\frac{65}{1386 n^5}-\frac{1549}{5544 n^7}+\frac{35}{128 n^8} \,,
\\
\frac{\langle Q_C \rangle_{nF_{5/2}}}{(Z\alpha)^8 m} =& \;
-\frac{4}{31185 n^3}+\frac{131}{99792 n^5}-\frac{23}{13860 n^7} \,,
\\
\frac{ \langle L_C \rangle_{nF_{5/2}}}{(Z\alpha)^8 m} =& \;
-\frac{281}{249480 n^3}+\frac{5975}{399168 n^5}-\frac{599}{13860 n^7} \,,
\\
\frac{E^{[8]}_d(nF_{5/2})}{(Z\alpha)^8 m} = & \;
-\frac{1033}{249480 n^3}+\frac{25219}{399168 n^5}
-\frac{8989}{27720 n^7} 
\nonumber\\
& \; +\frac{35}{128 n^8} \,,
\end{align}
\end{subequations}
For absolutely clarity, we should emphasize the 
all matrix elements are calculated 
with nonrelativistic Schr\"{o}dinger--Pauli
two-component reference-state wave functions
(Chap.~6 of Ref.~\cite{JeAd2022book}).
Summing up the results for the combined third-order 
perturbation theory term, and the mixed term,
one obtains
\begin{align}
\frac{E^{[8]}_c(nF_{5/2})}{(Z\alpha)^8 m} = & \;
-\frac{169721}{23950080 n^3}
-\frac{19}{1728 n^4}
+\frac{4560727}{41912640 n^5}
\nonumber\\
& \; +\frac{955}{3024 n^6}-\frac{1052}{693 n^7}+\frac{21}{16 n^8} \,,
\\[2ex]
\frac{E^{[8]}_m(nF_{5/2})}{(Z\alpha)^8 m} = & \;
\frac{262729}{23950080 n^3} +
\frac{5}{576 n^4} -
\frac{3652871}{20956320 n^5}
\nonumber\\
& \; -\frac{40}{189 n^6}+\frac{28271}{18480 n^7}-\frac{21}{16 n^8} \,,
\\[2ex]
\frac{E^{[8]}(nF_{5/2})}{(Z\alpha)^8 m} = & \;
-\frac{1}{3888 n^3}-\frac{1}{432 n^4} - \frac{1}{432 n^5}
+ \frac{5}{48 n^6}
\nonumber\\
& \; -\frac{5}{16 n^7}+\frac{35}{128 n^8} \,.
\end{align}
The latter terms confirm Eq.~\eqref{EDnF52}.

%
%
\subsection{Application: $\maybebm{F_{7/2}}$ States}
\label{sec3B}

The calculation proceeds in full analogy with
$nF_{5/2}$ states.
The Dirac--Coulomb energy finds the expansion,
\begin{multline}
\label{EDnF72}
E_\DD(nF_{7/2}) = m - \frac{(Z\alpha)^2 m}{2 n^2}
+ (Z\alpha)^4 m \left( \frac{3}{8 n^4} - \frac{1}{8 n^3} \right)  \\
- (Z\alpha)^6 m \left(
-\frac{1}{512 n^3}-\frac{3}{128 n^4}+\frac{3}{16 n^5}-\frac{5}{16 n^6}
\right)
\\
+ (Z\alpha)^8 m
\Bigl( 
-\frac{1}{16384 n^3}-\frac{3}{4096 n^4}-\frac{1}{1024 n^5}
\\
+\frac{15}{256 n^6}-\frac{15}{64 n^7}+\frac{35}{128 n^8}
\Bigr)
+ \calO(Z\alpha)^{10} \,.
\end{multline}
After lengthy algebra, we obtain the results,
\begin{subequations}
\label{IDnF72}
\begin{align}
\frac{\langle K_C \rangle_{nF_{7/2}}}{(Z\alpha)^8 m} =& \;
-\frac{2}{693 n^3}+\frac{65}{1386 n^5}-\frac{1549}{5544 n^7}+\frac{35}{128 n^8} \,,
\\
\frac{\langle Q_C \rangle_{nF_{7/2}}}{(Z\alpha)^8 m} =& \;
-\frac{5}{18144 n^3}+\frac{247}{90720 n^5}-\frac{1}{315 n^7} \,,
\\
\frac{ \langle L_C \rangle_{nF_{7/2}}}{(Z\alpha)^8 m} =& \;
\frac{13261}{7983360 n^3}-\frac{156853}{7983360 n^5}+\frac{9589}{221760 n^7} 
\,,
\\
\frac{E^{[8]}_d(nF_{7/2})}{(Z\alpha)^8 m} = & \;
-\frac{121}{80640 n^3}+\frac{2417}{80640 n^5}
-\frac{965}{4032 n^7}
\nonumber\\
& \; +\frac{35}{128 n^8} \,.
\end{align}
\end{subequations}

The combined third-order terms $E^{[8]}_c(nF_{7/2})$ and
mixed terms $E^{[8]}_m(nF_{7/2})$
find the representations,
\begin{align}
& \; \frac{E^{[8]}_c(nF_{7/2})}{(Z\alpha)^8 m} =
- \frac{548047}{232243200 n^3} -
\frac{253}{61440 n^4} 
\\
& \; \qquad 
+ \frac{2636603}{50803200 n^5} 
+ \frac{1427}{8064 n^6} -
\frac{22847}{20160 n^7}+\frac{21}{16 n^8} \,,
\nonumber\\[0.1133ex]
& \; \frac{E^{[8]}_m(nF_{7/2})}{(Z\alpha)^8 m} = 
\frac{55147}{14515200 n^3} +
\frac{13}{3840 n^4} -
\frac{8417851}{101606400 n^5} 
\nonumber\\
& \; \qquad - \frac{1909}{16128 n^6} +
\frac{7649}{6720 n^7} -
\frac{21}{16 n^8} \,.
\end{align}
The sum
\begin{multline}
\frac{E^{[8]}(nF_{7/2})}{(Z\alpha)^8 m} =
\frac{E^{[8]}_c(nF_{7/2})}{(Z\alpha)^8 m} +
\frac{E^{[8]}_m(nF_{7/2})}{(Z\alpha)^8 m} 
\\
+ \frac{E^{[8]}_d(nF_{7/2})}{(Z\alpha)^8 m} 
= -\frac{1}{16384 n^3}-\frac{3}{4096 n^4}-
\frac{1}{1024 n^5}
\\
+\frac{15}{256 n^6}
-\frac{15}{64 n^7}+\frac{35}{128 n^8} \,,
\end{multline}
reproduces the terms of order $(Z\alpha)^8$ from
the Dirac--Coulomb bound-state energy~\eqref{EDnF72}.

%
%
\section{Conclusions}
\label{sec4}

In this article, we have extended the 
treatment of the Foldy--Wouthuysen 
transformation to eighth order,
based on the scaling of the operators
outlined in Eq.~\eqref{scaling}.
The results are obtained 
by a straightforward application of the 
elimination of odd operators by repeated
unitary transformations of the 
form outlined in Eq.~\eqref{defSandU}.
The sixth-order terms 
[Eqs.~\eqref{calH6} and~\eqref{H6}]
are obtained in full agreement with the literature
(Refs.~\cite{ZaPa2010,HiLePaSo2013,PaYePa2016,HaZhKoKa2020,ZhMeQi2023}).
For the eighth-order terms, we give
results in Eqs.~\eqref{calH8} and~\eqref{H8}.
We have applied our general results
to the relativistic bound Coulomb problem in Sec.~\ref{sec3}.
An application to $nF_{5/2}$ and $nF_{7/2}$
states, which present large spin-orbit couplings,
confirms, analytically, that the 
Dirac--Coulomb bound-state energy 
can be obtained, within the Foldy--Wouthuysen
formalism, as a sum of combined third-order 
perturbative effects generated by the 
leading relativistic corrections [Eq.~\eqref{Ec}],
mixed fourth-order and sixth-order Hamiltonian terms [Eq.~\eqref{Em}],
and diagonal elements of the eighth-order Hamiltonian [Eq.~\eqref{Ed}].
The latter terms are obtained as diagonal elements
of our eighth-order terms for the Coulomb field,
evaluated on Schr\"{o}dinger--Pauli wave functions
(Chap.~6 of Ref.~\cite{JeAd2022book}).
As outlined in Ref.~\cite{HiLePaSo2013},
the results are important in a wider context, 
in view of the fact that the 
Foldy--Wouthuysen Hamiltonian determines (part of) 
the matching coefficients in the Hamiltonian
of Nonrelativistic Quantum Electrodynamics (NRQED).

%
%
\section*{Acknowledgements}

The author acknowledges helpful conversations with 
Gregory S.~Adkins.
This work has been supported by the National Science Foundation (Grant
PHY--2110294).

\appendix

%
%
\section{Comparison with the Literature}
\label{appa}

We compare our results with those of Ref.~\cite{ZhMeQi2023}.
One notes that the 
Foldy--Wouthuysen Hamiltonian derived in 
Ref.~\cite{ZhMeQi2023} is derived based on the 
Douglas--Kroll--Hess~\cite{DoKr1974,He1986} approach, which treats the 
kinetic-energy term in the relativistic 
Hamiltonian on a special footing and 
differs from the approach chosen here.
The result for the eighth-order terms given 
in Eq.~(8) of Ref.~\cite{ZhMeQi2023} can be written as
the sum $H'^{[8]} = \calK + \calD' + \calQ' + \calL'$, where
\begin{subequations}
\begin{align}
\calD' =& \; - \frac{\ii \, e^2}{32 m^4} \, 
[ \, \vec\sigma \cdot \vec E,
\vec\sigma \cdot \partial_t \vec E \, ] \,,
\\
\label{Qprime}
\calQ' =& \; -\frac{e^2}{32 m^5} \,  \;
\{ \, \vec\sigma \cdot \vec E,
2 \, (\vec\sigma \cdot \vec \pi)^2 \, \vec\sigma \cdot \vec E +
2 \, \vec\sigma \cdot \vec E (\vec\sigma \cdot \vec \pi)^2
\nonumber\\
& \; + \vec\sigma \cdot \vec \pi \, \vec\sigma \cdot \vec E \,
\vec\sigma \cdot \vec \pi \, \} \,, \\
\label{Lprime}
\calL' =& \; \frac{11 \ii e}{1024 m^6} [ (\vec\sigma \cdot \vec \pi)^4,
\{ \, \vec\sigma \cdot \vec \pi, \, \vec\sigma \cdot \vec E \, \} ]
\\
& \; - \frac{31 \ii e}{512 m^6} [ (\vec\sigma \cdot \vec \pi)^5,
\vec\sigma \cdot \vec E ]
- \frac{9 \, \ii \, e}{512 m^6} \,
[ \, (\vec\sigma \cdot \vec \pi)^3, \,
\nonumber\\
& \;
(\vec\sigma \cdot \vec \pi)^2 \, \vec\sigma \cdot \vec E
+ \vec\sigma \cdot \vec E \, (\vec\sigma \cdot \vec \pi)^2 +
\vec\sigma \cdot \vec \pi \, \vec\sigma \cdot \vec E \,
\vec\sigma \cdot \vec \pi \, ] \,.
\nonumber
\end{align}
\end{subequations}
Here, $\calD'$ is the term from
Eq.~(8) in Ref.~\cite{ZhMeQi2023}
which contains temporal derivatives, 
$\calQ'$ is the term from
Eq.~(8) in Ref.~\cite{ZhMeQi2023}
which is quadratic in the electric fields,
and $\calL'$ is the term from
Eq.~(8) in Ref.~\cite{ZhMeQi2023}
which is linear in the electric fields,
The kinetic term $\calK$ of eighth order 
agrees with the corresponding term from
Eq.~\eqref{calH8} here.

Specifically, the term $\calD'$
lacks the terms proportional to the 
anticommutator 
$\{  (\vec\sigma \cdot \vec \pi)^3, \vec\sigma \cdot \partial_t \vec E \}$
in comparison to our result for $\calD$ 
given in Eq.~\eqref{calH8}.
The results for $\calQ'$ and $\calL'$ differ from
those given in Eqs.~\eqref{calQalt} and~\eqref{calLalt} in the prefactors
of the individual terms.

One can easily specialize the 
4~operators given in Eqs.~\eqref{Qprime} and~\eqref{Lprime}
to the case of a Coulomb field, 
on the basis of Eq.~\eqref{scene_coul}.
This leads to the operators $Q'_C$ and $L'_C$.
The diagonal matrix elements, for $nF_{5/2}$ 
evaluate as follows,
\begin{subequations}
\begin{align}
\frac{\langle Q'_C \rangle_{nF_{5/2}}}{(Z\alpha)^8 m} =& \;
-\frac{37}{249480 n^3}+\frac{779}{498960 n^5}-\frac{29}{13860 n^7} \,,
\\
\frac{ \langle L'_C \rangle_{nF_{5/2}}}{(Z\alpha)^8 m} =& \;
-\frac{23}{20790 n^3}+\frac{1399}{95040 n^5}-\frac{593}{13860 n^7} \,.
\end{align}
For $nF_{7/2}$ states, we have the results,
\begin{align}
\frac{\langle Q'_C \rangle_{nF_{7/2}}}{(Z\alpha)^8 m} =& \;
-\frac{131}{332640 n^3}+\frac{1301}{332640 n^5}-\frac{16}{3465 n^7} \,,
\\
\frac{ \langle L'_C \rangle_{nF_{7/2}}}{(Z\alpha)^8 m} =& \;
\frac{947}{532224 n^3}-\frac{7921}{380160 n^5}+\frac{1101}{24640 n^7} \,.
\end{align}
\end{subequations}
These results differ individually
from those given in Eqs.~\eqref{IDnF52}
and~\eqref{IDnF72},
for the diagonal matrix elements of the 
operators $Q_C$ and $L_C$, but their sum 
reproduces our results for both
fine-structure components ($F_{5/2}$ and 
$F_{7/2}$) investigated here.
We have carried out similar calculations
for states with a different angular symmetry
(e.g., $G$ states) and find a similar behavior.
These observations support the conjecture
that the eighth-order Hamiltonian derived 
here and in Ref.~\cite{ZhMeQi2023}
lead to equivalent diagonal elements
for hydrogenic reference states.

However, the Hamiltonians derived 
here and in Ref.~\cite{ZhMeQi2023} are 
not equivalent for time-dependent
problems. 
Let us consider a binding Coulomb field 
added to an external plane-wave laser field
(in the length gauge), polarized along the $z$ axis, 
with
\begin{equation}
e \, A^0 = -\frac{Z\alpha}{r} - e \frac{E_L}{\omega_L} z \, \sin(\omega_L \, t)\,,
\end{equation}
where $E_L$ is the peak laser field during a 
laser period, and $\omega_L$ is the laser angular
frequency. The vector potential still vanishes,
so that the relation $\vec\pi = \vec p$ is retained.
The spatially homogeneous, but time-dependent, laser field is
$\vec E_L(t) = \hat{\mathrm{e}}_z \, E_L \, \cos(\omega_L t)$.
The total $\vec E$ field (Coulomb$+$laser field) fulfills
$e \vec E = 
\vec\nabla \frac{Z\alpha}{r} + e \hat{\mathrm{e}}_z  
E_L \, \cos(\omega_L \, t)$.
The commutator $[ \, \vec\sigma \cdot \vec E,
\vec\sigma \cdot \partial_t \vec E \, ]$ vanishes,
but the term
\begin{equation}
H^{[8]} \sim \calD \sim \frac{e}{48 m^5} \,
\{  (\vec\sigma \cdot \vec p)^3, \vec\sigma \cdot \partial_t \vec E_L(t) \} \,,
\end{equation}
from the $\calD$ term in our Eq.~\eqref{H8},
generates a contribution proportional to 
$\sin(\omega_L \, t)$, in view of the 
time derivative of the laser field.
Its sinusoidal (as opposed to cosinusoidal) 
time dependence cannot be compensated
by any term proportional to the laser field
itself, $ \vec E_L(t) \propto
\cos(\omega_L t)$, i.e., the sinusoidal term cannot be compensated
by any other term in $H^{[8]}$ which is
free from time derivatives of the electric
field. Hence, the Hamiltonians derived 
here and in Ref.~\cite{ZhMeQi2023} cannot 
be completely equivalent for time-dependent problems. 

The eighth-order Hamiltonians derived
here and in Ref.~\cite{ZhMeQi2023} could potentially 
be equivalent up to a unitary transformation,
in a somewhat distant analogy to the 
unitary transformation given in Eq.~(19)
of Ref.~\cite{PaYePa2016}, which was applied
in Ref.~\cite{PaYePa2016}
to different forms of the sixth-order Foldy--Wouthuysen
Hamiltonian (see also the nonstandard Foldy--Wouthuysen
transformation used in Refs.~\cite{Pa2005,ZhMeQi2019}).
A potential unitary transformation 
which brings the results 
communicated in Ref.~\cite{ZhMeQi2023} and
those derived here into agreement,
would only need to affect the eighth-order terms
because the sixth-order terms indicated here 
and in Ref.~\cite{ZhMeQi2023} are identical.
The special form of the $S$ operator,
which generates the Foldy--Wouthuysen $U$
transformation via the relation $U = \exp(\ii S)$
for the nonstandard approach from Refs.~\cite{Pa2005,ZhMeQi2019},
has recently been highlighted
in Eq.~(17) of Ref.~\cite{ToMeZhQi2023}.
In general, when two Hamiltonians are related by 
a unitary transformation, then their matrix 
elements are identical {\em provided} 
one also applies the unitary transformation
to the wave functions. Within this context,
we mention that unitary (gauge) transformations
of the wave functions can change
their physical interpretation.
This (perhaps surprising) fact is 
relevant for the quantum dynamical formulation
of laser-induced processes off resonance
(see the footnote on p.~268 of
Ref.~\cite{La1952} and 
the eludicating discussion in Ref.~\cite{ScBeBeSc1984}).

Another indication that the Hamiltonians
derived here and in Ref.~\cite{ZhMeQi2023} cannot 
be completely equivalent,
stems from the calculation of 
off-diagonal matrix elements of the
Hamiltonian. As an example,
we shall calculate off-diagonal elements
of the operators $Q_C$, $Q'_C$, $L_C$ and $L'_C$,
sandwiched between $|4F_{5/2}\rangle$
and $|6F_{5/2}\rangle$ states,
with the results
\begin{subequations}
\begin{align}
\frac{\langle 4F_{5/2} | Q_C | 6F_{5/2} \rangle}{ (Z \alpha)^8 m }  = & \;
- \frac{761 }{1\,417\,500\,000} \,,
\\[0.1133ex]
\frac{\langle 4F_{5/2} | Q'_C | 6F_{5/2} \rangle}{ (Z \alpha)^8 m } = & \;
- \frac{23 }{40\,500\,000} \,,
\\[0.1133ex]
\frac{\langle 4F_{5/2} | L_C | 6F_{5/2} \rangle}{ (Z \alpha)^8 m } = & \;
- \frac{5\,007\,493 }{1\,134\,000\,000\,000}\,,
\\[0.1133ex]
\frac{\langle 4F_{5/2} | L'_C | 6F_{5/2} \rangle}{ (Z \alpha)^8 m } = & \;
- \frac{1\,627\,151 }{378\,000\,000\,000} \,.
\end{align}
The sum of these terms is
\begin{align}
\frac{ \langle 4F_{5/2} | Q_C + L_C | 6F_{5/2} \rangle 
}{ (Z \alpha)^8 m } = & \;
- \frac{5\,616\,293}{1\,134\,000\,000\,000} \,,
\\[0.1133ex]
\frac{\langle 4F_{5/2} | Q'_C + L'_C | 6F_{5/2} \rangle 
}{ (Z \alpha)^8 m } = & \;
- \frac{5\,525\,453 }{1\,134\,000\,000\,000} \,.
\end{align}
\end{subequations}
Numerically, the difference between 
$\langle 4F_{5/2} | Q_C + L_C | 6F_{5/2} \rangle = 
-4.952 \times 10^{-6} \, (Z \alpha)^8 m$ and
$\langle 4F_{5/2} | Q'_C + L'_C | 6F_{5/2} \rangle = 
-4.873 \times 10^{-6} \, (Z \alpha)^8 m$ is about $1.6 \,$\%.
Because the kinetic terms 
derived here and in Ref.~\cite{ZhMeQi2023} agree,
this observation implies that the off-diagonal matrix elements
derived from the total $H^{[8]}$ and $H'^{[8]}$ differ.

For off-diagonal matrix elements of $|4F_{7/2}\rangle$
and $|6F_{7/2}\rangle$ states,
the following results are obtained,
\begin{subequations}
\begin{align}
\frac{ \langle 4F_{7/2} | Q_C | 6F_{7/2} \rangle
}{ (Z \alpha)^8 m } = & \;
- \frac{3587}{2\,835\,000\,000} \,,
\\[1.1133ex]
\frac{ \langle 4F_{7/2} | Q'_C | 6F_{7/2} \rangle 
}{ (Z \alpha)^8 m } = & \;
- \frac{49 }{27\,000\,000} \,,
\\[1.1133ex]
\frac{ \langle 4F_{7/2} | L_C | 6F_{7/2} \rangle 
}{ (Z \alpha)^8 m } = & \;
\frac{14\,493\,239 }{1\,134\,000\,000\,000}\,,
\\[1.1133ex]
\frac{ \langle 4F_{7/2} | L'_C | 6F_{7/2} \rangle 
}{ (Z \alpha)^8 m } = & \;
\frac{1\,731\,391 }{252\,000\,000\,000} \,.
\end{align}
The sum of these terms is
\begin{align}
\frac{ \langle 4F_{7/2} | Q_C + L_C | 6F_{7/2} \rangle 
}{ (Z \alpha)^8 m } = & \;
\frac{11\,623\,639}{2\,268\,000\,000\,000} \,,
\\[0.1133ex]
\frac{ \langle 4F_{7/2} | Q'_C + L'_C | 6F_{7/2} \rangle =
}{ (Z \alpha)^8 m } = & \;
\frac{3\,822\,173 }{756\,000\,000\,000} \,.
\end{align}
\end{subequations}
One observes that, numerically, the difference between
$\langle 4F_{7/2} | Q_C + L_C | 6F_{7/2} \rangle =
5.125  \times 10^{-6} \, (Z \alpha)^8 m$ and
$\langle 4F_{7/2} | Q'_C + L'_C | 6F_{7/2} \rangle =
5.056 \times 10^{-6} \, (Z \alpha)^8 m$ is about $1.4 \,$\%.

In the very recent paper~\cite{ToMeZhQi2023},
the standard approach to the Foldy--Wouthuysen
transformation is applied to 
obtain a result for the eighth-order terms
communicated in Eq.~(16) of Ref.~\cite{ToMeZhQi2023}.
The result from Ref.~\cite{ToMeZhQi2023}
differs from our result, given in Eq.~\eqref{H8},
in the sign of the term
$\frac{e}{48 m^5} \, \beta \,
\{ \, (\vec\Sigma \cdot \vec \pi)^3, 
\vec\Sigma \cdot \partial_t \vec E \, \}$.
In view of the aspects discussed 
in this Appendix, we leave the final clarification of the 
eighth-order terms derived here  to those 
communicated in 
Refs.~\cite{ZhMeQi2019,ZhMeQi2023,ToMeZhQi2023} as an open 
problem for future
investigations.

\end{document}